\documentclass{article}
\usepackage{spconf,amsmath,amssymb,graphicx,hyperref}
\usepackage{booktabs}     
\usepackage{float}
\usepackage{xcolor}       
\usepackage{algorithm}    
\usepackage{algorithmic}
\usepackage{amsmath}
\usepackage{mathtools}
\usepackage{orcidlink}

\title{Model-Guided Design of Low-Context Speech Probes for Cochlear Synaptopathy }
\name{%
  Ahsan Jamal Cheema$^{1,2}$\,\orcidlink{0000-0002-5384-0973}
  \quad
  David Meng$^{3}$\,\orcidlink{0000-0002-8830-2606}
  \quad
  Jorge Mejia$^{3}$\,\orcidlink{0000-0002-9624-2842}
  \quad
  Sanna Hou$^{3}$\,\orcidlink{0000-0001-9956-8963}
  \quad
  Sunil Puria$^{1,2}$\,\orcidlink{0000-0002-2810-3541}%
  \thanks{Thanks to US NIH-NIDCD (Grant \# R01DC007910) and The Government of Australia Department of Health, Disability and Ageing for funding.}
}
\address{$^{1}$ Harvard University, Cambridge MA, USA\\
$^{2}$ Eaton-Peabody Laboratories, Massachusetts Eye and Ear (MEEI), Boston MA, USA\\
$^{3}$ National Acoustic Laboratories (NAL), Sydney, Australia}

\begin{document}
\ninept
\maketitle
\begin{abstract}
Cochlear neural degeneration (CND) can impair suprathreshold coding without elevating pure-tone thresholds, complicating its diagnosis when it coexists with hair cell loss. We present a unified comparison of temporal and noise-based probes for CND detection using low-context vowel–consonant–vowel (VCV) syllables to reduce linguistic and contextual cues. Using a phenomenological auditory nerve model, we simulated responses to 21 VCV tokens under time compression, reverberation, and speech-in-noise conditions across presentation levels and seven CND profiles. We computed mutual information (MI) between inner hair cell potentials and auditory nerve neurograms and quantified information loss relative to a normal-hearing baseline. Time compression and amplitude-modulated (AM) noise produced the largest modeled information losses. We then evaluated these stimuli in a consonant-identification study involving 36 listeners with normal audiograms, 12 of whom reported difficulty understanding speech in noise. Neither 40\% time compression in quiet nor AM noise alone distinguished listeners with and without these difficulties. However, compressed speech presented in AM noise separated the two groups. This partial agreement between model predictions and behavior supports our MI-based stimulus design framework and motivates further evaluation of combined temporal and noise-based probes for CND detection.
\end{abstract}
\section{Introduction}

Cochlear neural degeneration (CND), including the loss of synapses between inner hair cells (IHCs) and auditory nerve fibers (ANFs), can impair suprathreshold coding without elevating audiometric thresholds~\cite{Kujawa2009-ie, Liberman2017-pf, Hickox2014Synaptopathy, Bramhall2019MissionImpossible, oxenham2016predicting}. Animal studies indicate preferential vulnerability of low- and medium-spontaneous-rate (SR) fibers, which have relatively wide dynamic ranges and are thought to contribute to the encoding of suprathreshold signals in noise~\cite{Furman2013-mz, Bharadwaj2014-us}. CND has therefore been proposed as one explanation for why listeners with normal audiograms report difficulty understanding speech in noisy environments~\cite{DiNino2022-xm}. However, detecting CND noninvasively remains challenging, particularly when it coexists with hair cell loss~\cite{Liberman2017-pf}. Hair cell dysfunction reduces the input driving ANF responses, whereas CND reduces the number of fibers carrying that information. Both can reduce neural output, making their contributions difficult to distinguish using psychophysical measures~\cite{plack2016toward}. It is therefore important to identify stimulus conditions that are sensitive to CND and can distinguish its effects from those of hair cell dysfunction.
Temporal and noise-based manipulations have been used to investigate these deficits. Temporal manipulations, such as time compression and reverberation, alter the timing and temporal structure of speech~\cite{Liberman2016-ok, Cheema2025-qw}. Given differences in ANF dynamic range, saturation, and recovery from forward masking, these manipulations may place greater demands on the remaining fiber population. Noise-based probes are motivated by the hypothesis that background noise drives high-SR fibers toward saturation, increasing the importance of the low- and medium-SR fibers that are vulnerable to CND~\cite{DiNino2022-xm}. Comparing these manipulations within a common quantitative framework may help identify which conditions are most sensitive to neural loss.
In our previous work~\cite{Cheema2025-qw}, we developed a mutual information (MI) framework to quantify information loss at the IHC--ANF junction. Using consonant--vowel--consonant (CVC) words from the NU-6 corpus, we found that time compression produced the largest modeled information loss among the conditions tested. However, these findings need to be evaluated using stimuli with different acoustic and linguistic properties. Speech recognition depends on both envelope and temporal fine structure (TFS) cues, whose contributions vary with the listening condition~\cite{Smith2002-yf, lorenzi2006, hopkins2010}. Furthermore, recognition of meaningful words can benefit from lexical knowledge and cognitive processes that compensate for degraded sensory input~\cite{Akeroyd2008-yx}, potentially reducing sensitivity to peripheral deficits~\cite{plack2016toward}.

In this study, we extend the MI framework to vowel--consonant--vowel (VCV) nonsense syllables, which reduce lexical cues while providing sustained vowel segments and vowel--consonant transitions. We compare temporal manipulations, including time compression, reverberation, and their combination, with stationary and amplitude-modulated noise using the same corpus across four presentation levels and seven CND profiles. We then evaluate the two conditions ranked highest by the model, individually and in combination, in a consonant-identification study involving listeners with normal audiograms, with and without self-reported difficulty understanding speech in noise. Hearing difficulties in these listeners may reflect suprathreshold coding deficits associated with CND that are not captured by audiometric thresholds. We ask whether these model-selected conditions distinguish the two groups, providing a behavioural test of our MI-based approach to designing stimuli sensitive to deficits potentially associated with CND.

\vspace{-10 pt}
\section{Methods}

\subsection{Speech Corpus and Probe Conditions}

We recorded 21 vowel--consonant--vowel (VCV) nonsense syllables in a sound-proof chamber at a sampling rate of 44.1~kHz. The corpus comprised eight consonants (/p, t, k, m, n, s and l/) presented in three vowel contexts (/aCa/, /iCi/, and /oCo/). Nonsense syllables were used to reduce contributions from lexical knowledge and semantic prediction~\cite{Dubno1982-iv}. Each token was processed under temporal and noise-based conditions. The temporal conditions were presented without added noise and included (1) unprocessed speech, (2) 40\% time compression, (3) reverberation with $T_{60} \approx 0.3$~s, (4) 40\% compression with reverberation, (5) 65\% compression, and (6) 65\% compression with reverberation. Throughout this study, $x$\% compression indicates that the token duration was reduced to $x$\% of its original duration. Thus, 40\% compression represents a greater reduction in duration than 65\% compression. Time compression was implemented using a pitch-preserving algorithm from the synchronous overlap-add (SOLA) family. The noise conditions included stationary noise shaped to the long-term average speech spectrum and amplitude-modulated (AM) noise, both presented at 0~dB SNR. AM noise consisted of speech-shaped noise with sinusoidal amplitude modulation at 4 Hz and 100\% modulation depth. For the simulations, stimuli were level-normalized and presented at 65, 80, 90, and 95~dB SPL.
\subsection{Auditory Nerve Model and CND Profiles}
We used a phenomenological auditory nerve model~\cite{Bruce2018-ma, Guest2024-vp} to simulate responses at 50 logarithmically spaced characteristic frequencies (CFs) between 125~Hz and 16~kHz. At each CF, the model provided the IHC receptor potential, $V_{\mathrm{IHC}}(f,t)$, and a fine-timing ANF neurogram, $N(f,t)$, expressed in spikes/s. Both outputs were sampled at 44.1~kHz. Hair cell dysfunction was simulated using a sloping audiogram with thresholds of 0, 0, 10, 20, 23, 45, and 75~dB~HL at 0.125, 0.25, 0.5, 1, 2, 4, and 8~kHz, respectively. Two-thirds of the threshold elevation was assigned to outer hair cell dysfunction and one-third to IHC dysfunction~\cite{moore2004}. CND was simulated by varying the number of low-, medium-, and high-SR fibers per IHC. The intact population comprised 22 fibers per IHC, with [LS, MS, HS] = [5,5,12]. We evaluated seven ANF profiles, including this intact population and six profiles with neural loss. Removing 40\%, 80\%, and 100\% of the LS and MS fibers produced populations of [3,3,12], [1,1,12], and [0,0,12], respectively. Three additional profiles combined complete LS and MS loss with approximately 20\%, 40\%, and 60\% HS loss, giving [0,0,10], [0,0,7], and [0,0,5]. This ordering reflected the preferential vulnerability of low- and medium-SR fibers reported in animal studies~\cite{Furman2013-mz}.
\vspace{-10 pt}
\subsection{Mutual Information Analysis}
We quantified information transmission at the IHC--ANF junction using the mutual information (MI) between $V_{\mathrm{IHC}}$ and $N$ at each CF~\cite{Cheema2025-qw}. MI was estimated from a joint histogram with $B = 1024$ bins, selected using the bias--variance analysis described previously~\cite{Cheema2025-qw}:
\begin{equation}
I_f =
\sum_{i,j}
\hat{p}(v_i,n_j)
\log_2
\left[
\frac{\hat{p}(v_i,n_j)}
{\hat{p}(v_i)\hat{p}(n_j)}
\right],
\end{equation}
where $\hat{p}(v_i,n_j)$ denotes the estimated joint probability and $\hat{p}(v_i)$ and $\hat{p}(n_j)$ denote the corresponding marginal probabilities.
MI was summarized across CFs using
\begin{equation}
\mathrm{AUC} =  \sum_f I_f \log(f_c) \cdot\Delta(\log f_c).
\end{equation}
For each probe and presentation level, we simulated a normal-hearing (NH) reference without CND and responses for each combination of hair cell dysfunction and ANF profile. Information loss for profile $k$ was calculated relative to the NH reference:
\begin{equation}
\Delta\mathrm{AUC}^{(k)} = \mathrm{AUC}^{\mathrm{NH}} - \mathrm{AUC}^{(k)}.
\end{equation}

\subsection{Behavioural Validation Study}
\label{sec:behav_methods}
Thirty-six adults with "clinically normal" pure-tone thresholds $\leq 20$~dB~HL at 0.25--8~kHz in both ears were recruited for the study approved as per the the Hearing Australia Human Research Ethics Committee (Approval \# HAHREC2026-25). Participants who did not report any hearing in noise difficulties were assigned to a Control group ($n = 24$; 12 female; age $26.1 \pm 5.8$~years) and a self-reported hearing difficulty (SRHD) group ($n = 12$; 8 female; age $29.2 \pm 7.3$~years) using the Revised Hearing Handicap Inventory--Screening (RHHI-S) score criteria $\geq$ 6. The SRHD group allowed us to assess performance in listeners with difficulties potentially associated with CND. Compared with the Control group, the SRHD group had higher RHHI-S scores ($23.0 \pm 11.1$ vs.\ $0.8 \pm 1.6$) and lower SSQ scores ($5.0 \pm 1.3$ vs.\ $7.7 \pm 0.9$). Mean four-frequency pure-tone averages were $5.8 \pm 3.1$ and $5.3 \pm 3.7$~dB~HL in the SRHD and Control groups, respectively. Extended high-frequency thresholds at 9--16~kHz were $12.2 \pm 13.6$ and $7.2 \pm 12.9$~dB~HL, respectively ($p = .32$). All participants achieved 100\% sentence recognition in quiet.

\begin{figure}[t]
\centering
\includegraphics[width=0.9\columnwidth]{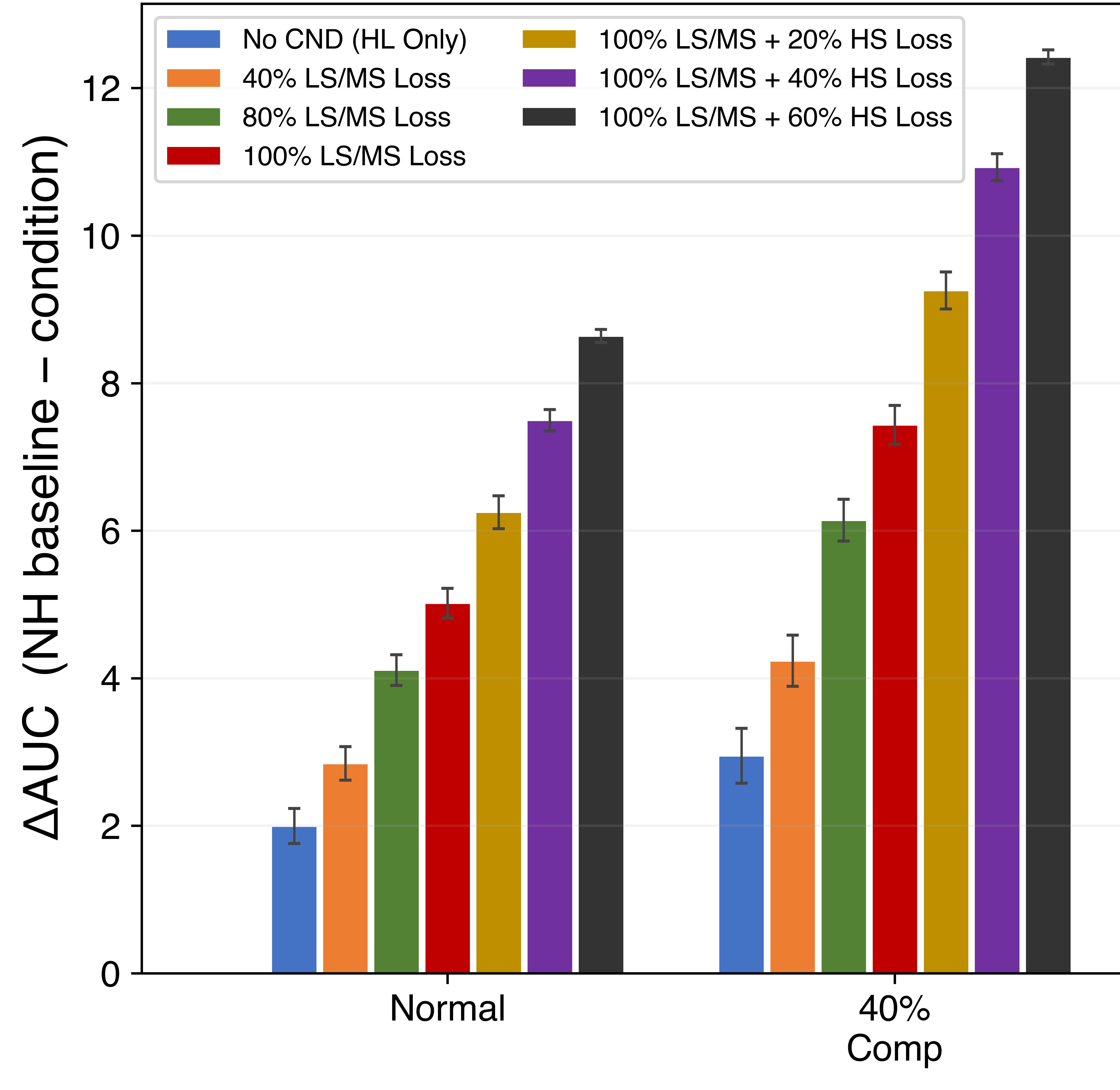}
\caption{$\Delta$AUC ($\text{AUC}^{(\text{NH})} - \text{AUC}^{(k)}$, bits$\cdot$log(Hz)) for the $V_\text{IHC} \to \text{AN}$ pathway at 95~dB SPL for unprocessed and 40\%-compressed VCV tokens across seven CND profiles. 40\% compression produced the largest information loss of all six temporal probes (Sec.~3.1).}
\label{fig:clean_delta}
\end{figure}

\begin{figure*}[t]
\centering
\includegraphics[width=0.84\textwidth]{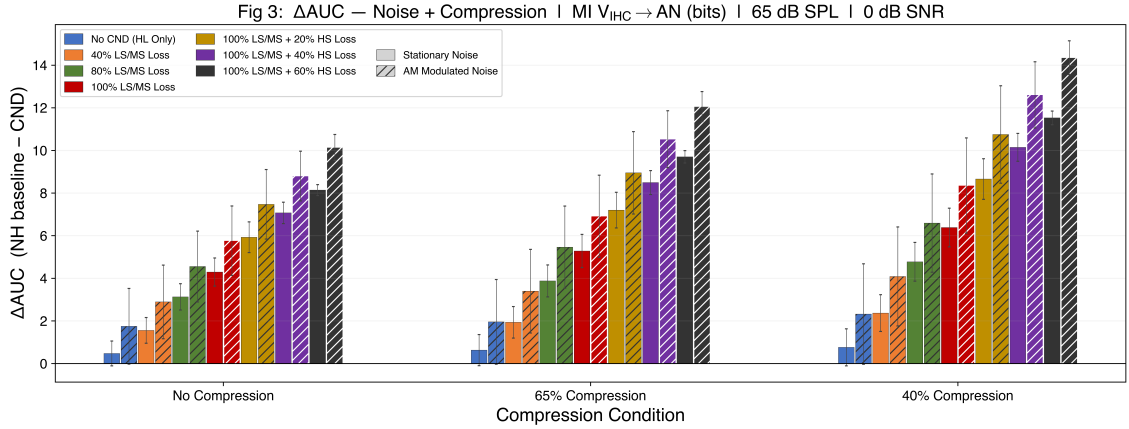}
\caption{$\Delta$AUC for the $V_\text{IHC} \to \text{AN}$ pathway at 65~dB SPL and 0~dB SNR for noise-overlaid VCV tokens without compression (left), with 65\% (center) and with 40\% compression (right). Solid bars: stationary noise; hatched bars: AM noise. Compressing the noise-overlaid signal progressively increases CND-induced information loss, and 40\% compression in AM noise produces the largest $\Delta$AUC in this study.}
\label{fig:mix_delta}
\end{figure*}

The behavioural conditions were selected from the model comparisons of temporal and noise-based probes (Figs.~\ref{fig:clean_delta} and \ref{fig:mix_delta}). We tested 40\% time compression and AM noise, both individually and in combination. This produced a $2 \times 2$ within-subject design comprising unprocessed and 40\%-compressed VCV tokens presented in quiet or AM noise. The behavioural corpus comprised 21 tokens (7 consonants $\times$ 3 vowels). The behavioural noise conditions differed from the simulations in SNR. Whereas the simulations used 0~dB SNR, behavioural testing was conducted at each listener's speech reception threshold for 50\% correct identification (SRT${50}$) for stimuli presented in noise, which was $\approx -16$~dB SNR for both groups. Individual SRT$_{50}$ values were measured for unprocessed VCV tokens presented in AM noise using an adaptive procedure. After an initial staircase phase, thresholds were estimated by fitting a logistic psychometric function to the accumulated response data, and the stimulus SNR was adjusted iteratively to converge on 50\% correct recognition. Both unprocessed and compressed tokens were then presented at the same individual SNR. This calibration was used because model MI does not directly specify the SNR required for a given level of behavioural performance. Testing at individual SRT$_{50}$ values targeted comparable baseline performance for unprocessed speech in AM noise, allowing us to assess the additional effect of compression relative to this baseline. Additional simulations showed that reducing SNR below 0~dB produced little change in $\Delta\mathrm{AUC}$ (Sec.~\ref{sec:noise_results}).

Stimuli were delivered through Sennheiser (model: HD215) headphones at 65~dB SPL. Participants identified the medial consonant from a closed set of options by clicking the corresponding item on the screen using a computer mouse. Reaction time was measured relative to stimulus onset. Each condition comprised 21 trials. The unprocessed and 40\%-compressed VCV tokens were randomly intermixed to create a block of 42 trials. The block was presented twice in the presence of AM noise and once in quiet. The presentation order of the quiet and noise conditions was counterbalanced across participants. SRT$_{50}$ values were compared between groups using Welch's $t$-test. Identification accuracy and mean reaction time were analyzed using linear mixed-effects models. Group, Listening condition (Quiet, AM noise), and Speech condition (Unprocessed, Compressed) were included as fixed effects, together with all interactions. Participant was included as a random intercept. Estimated marginal means and planned comparisons were used to evaluate condition and group differences, with $\alpha = .05$.

\vspace{-10 pt}
\section{Results}

\subsection{Temporal Probes in Quiet}
Figure~\ref{fig:clean_delta} shows $\Delta$AUC for unprocessed and 40\%-compressed speech at 95~dB SPL, which produced the largest information loss across the presentation levels tested (65--95~dB SPL). For all temporal conditions, $\Delta$AUC increased monotonically with CND severity. At the most severe CND profile, 40\% compression produced the largest $\Delta$AUC, approximately 12.5~bits$\cdot$log(Hz), compared with 10.5~bits$\cdot$log(Hz) for 65\% compression. Adding reverberation reduced these values to approximately 10.7 and 9.1~bits$\cdot$log(Hz), respectively. Reverberation alone produced the smallest $\Delta$AUC, approximately 7.8~bits$\cdot$log(Hz). These results identify time compression as the temporal manipulation producing the greatest modeled information loss, consistent with our previous findings~\cite{Cheema2025-qw, Cheema2025NSIM}.
\vspace{-10 pt}
\subsection{Noise Type and Time Compression}
\label{sec:noise_results}
Figure~\ref{fig:mix_delta} shows $\Delta$AUC for unprocessed, 65\%-compressed, and 40\%-compressed speech in stationary and AM noise at 65~dB SPL and 0~dB SNR. Even without compression, the noise conditions produced information losses comparable in magnitude to those observed for 40\% compression in quiet. AM noise consistently produced larger $\Delta$AUC than stationary noise across CND profiles, reaching approximately 10 and 8.5~bits$\cdot$log(Hz), respectively, at the most severe profile. Time compression further increased $\Delta$AUC for both noise types. In AM noise, $\Delta$AUC reached approximately 12~bits$\cdot$log(Hz) for 65\% compression and 14.5~bits$\cdot$log(Hz) for 40\% compression. The combination of 40\% compression and AM noise therefore produced the largest modeled information loss among the conditions tested. For the noise conditions, increasing the presentation level above 65~dB SPL or reducing the SNR below 0~dB produced no appreciable further increase in $\Delta$AUC.
\vspace{-12 pt}
\subsection{Behavioural Validation}
\label{sec:behav_results}
\begin{figure}[t]
\centering
\includegraphics[width=0.95\columnwidth]{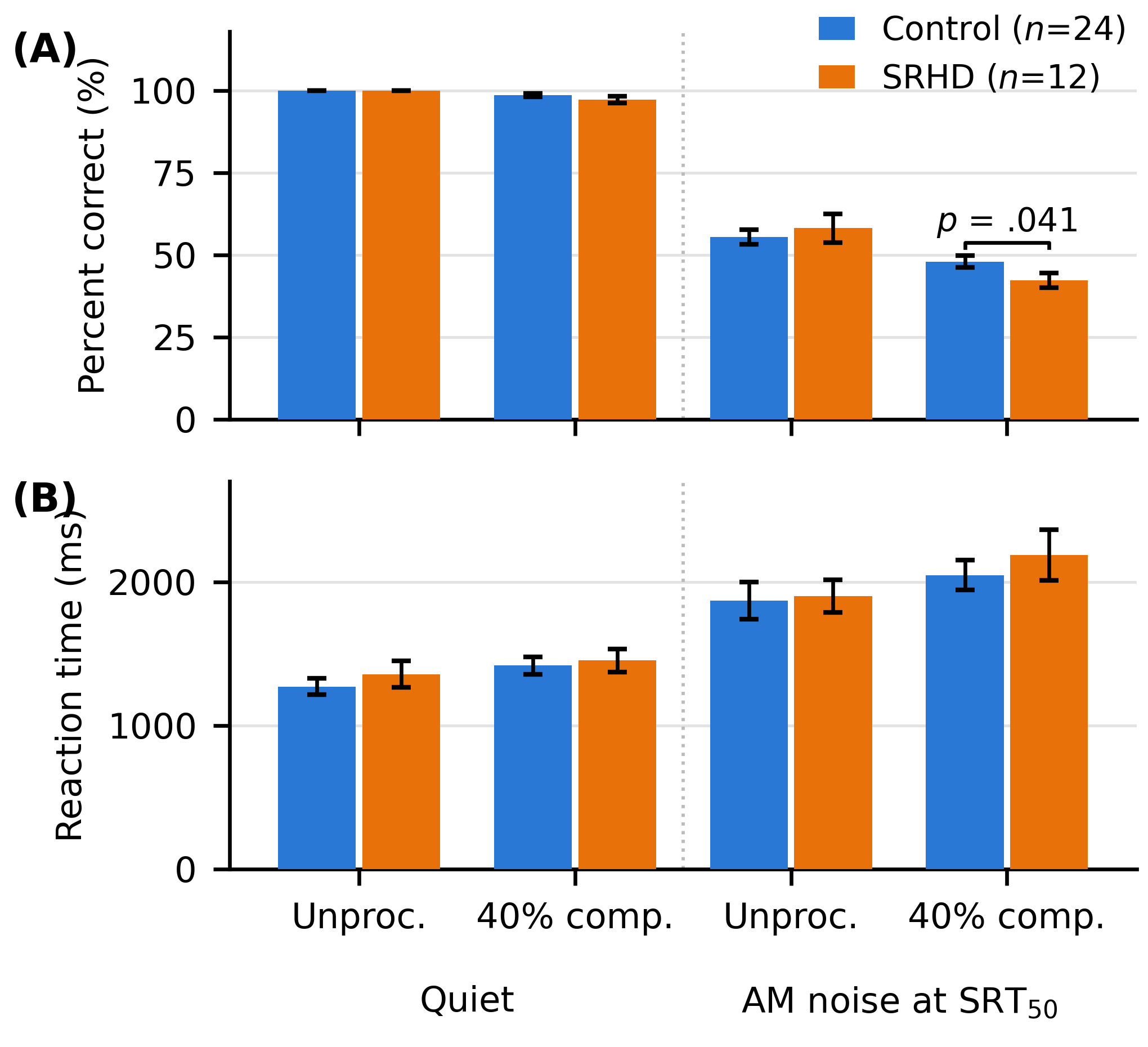}
\caption{Behavioural results: (A) consonant identification accuracy and (B) mean reaction time for the Control and self-reported hearing difficulty (SRHD) groups, for unprocessed and 40\%-compressed VCV tokens in quiet and in AM noise at each listener's SRT$_{50}$. Bars: group means; error bars: $\pm 1$ SEM. The groups differ only for compressed speech in AM noise (planned comparison, $p = .041$).}
\label{fig:behav}
\end{figure}

Figure~\ref{fig:behav} summarizes identification accuracy and reaction time for unprocessed and 40\%-compressed speech in quiet and AM noise. The behavioural noise conditions were presented at each listener's SRT$_{50}$. SRT$_{50}$ did not differ significantly between groups. Mean thresholds were $-16.5 \pm 0.5$~dB SNR (mean $\pm$ SEM) for Controls and $-15.5 \pm 1.3$~dB SNR (mean $\pm$ SEM)  for the SRHD group (Welch $t(14.6) = -0.67$, $p = .52$, $d = -0.28$). Thus, behavioural testing in noise was conducted approximately 16~dB below the SNR used in the primary simulations.
In quiet, both groups achieved 100\% identification accuracy for unprocessed tokens. Compression reduced accuracy by 2.1 percentage points overall, but this effect was not significant ($p = .22$). In AM noise, compression reduced accuracy by 11.7 percentage points. The mixed-effects model showed significant main effects of Listening condition and Speech condition, together with a Listening $\times$ Speech interaction (all $p < .001$). There was no significant main effect of Group ($F(1,34) = 0.38$, $p = .54$). We also found a significant Group $\times$ Speech interaction ($F(1,102) = 4.06$, $p = .047$). Averaged across listening conditions, the compression penalty was approximately twice as large in the SRHD group as in Controls: 9.3 percentage points ($p < .001$) compared with 4.5 percentage points ($p = .002$). Planned comparisons showed no significant group difference for unprocessed speech in AM noise (Control: 55.5\%; SRHD: 58.1\%; $p = .33$). For compressed speech in AM noise, however, accuracy was significantly lower in the SRHD group than in Controls (42.3\% vs.\ 47.9\%; $p = .041$). A significant group difference was therefore observed only when time compression and AM noise were combined (Fig. \ref{fig:behav} A).
Reaction times also increased with task difficulty (Fig.~\ref{fig:behav} B). The model showed significant main effects of Listening condition ($F(1,102) = 145.2$, $p < .001$) and Speech condition ($F(1,102) = 11.5$, $p = .001$). Compression increased reaction time by 232~ms in AM noise ($p = .004$), whereas its effect in quiet was not significant ($p = .11$). Neither the main effect of Group ($p = .58$) nor any interaction involving Group was significant (all $p > .44$). For compressed speech in AM noise, mean reaction times were numerically longer in the SRHD group than in Controls (2187 vs.\ 2047~ms), but this difference was not significant ($p = .38$) Fig. \ref{fig:behav} B).
 
\vspace{-10 pt}
\section{Discussion}
\vspace{-12 pt}
\subsection{Effects of Time Compression and AM Noise}
Time compression and AM noise may increase demands on neural coding through complementary mechanisms. Compression shortens speech cues and the intervals between successive events, while AM noise limits their availability through alternating periods of masking. These manipulations may interact with synaptic adaptation and recovery processes represented in the auditory nerve model~\cite{Bruce2018-ma}, increasing the effect of neural loss on the population response. For the noise conditions, increasing presentation level above 65~dB SPL or reducing SNR below 0~dB produced little additional increase in $\Delta$AUC, suggesting that higher levels were unnecessary to obtain a large modeled contrast within the conditions tested.
\vspace{-10 pt}
\subsection{Behavioural Support for the MI Framework}
The model guided the selection of conditions for behavioural testing. Neither compression in quiet nor unprocessed speech in AM noise significantly separated the groups, whereas compressed speech in AM noise did. The larger compression penalty in the SRHD group provides partial support for using modeled information loss to identify informative stimulus combinations.
This difference emerged at individual SRT${50}$ values, approximately 16~dB below the SNR used in the primary simulations. The finding is consistent with the limited SNR dependence of the modeled contrast, although the complete stimulus ranking was not tested behaviourally. Similar audiometric thresholds and the use of nonsense syllables limited some audiometric and linguistic confounds. However, these controls and the SRT${50}$ calibration did not eliminate central contributions or establish CND as the cause of the group difference. Reaction-time differences were not significant.
\vspace{-12 pt}
\subsection{Information Loss and Performance in Quiet}
The difference between model predictions and behaviour in quiet may partly reflect the near-ceiling performance of the closed-set task. Both groups achieved 100\% accuracy for unprocessed speech and remained above 97\% after compression, suggesting that the remaining cues were sufficient for consonant identification. Although $\Delta$AUC quantifies modeled information loss, it does not directly specify how much information the task actually requires. Compression shortens consonant cues, but these cues may remain accessible in quiet. In AM noise, their availability also depends on their timing relative to the masker, and shorter consonants may span fewer dips, reducing opportunities to recover informative speech glimpses~\cite{Cooke2006-gl}. Masker-driven adaptation may further affect responses to speech during subsequent dips, with differences in recovery across ANF types providing a possible mechanism for greater effects of neural loss under these conditions~\cite{Harris1979-hd, Relkin1991-rd}. Perceptual adaptation may also help listeners use the remaining cues in compressed speech~\cite{Dupoux1997-dg}, but such processes are not represented by the peripheral model. The framework can therefore guide the selection of potentially informative stimuli, while the behavioural consequences of modeled information loss also depend on task demands and the listener's ability to use the remaining cues i.e., decoder efficiency.

\vspace{-10 pt}
\section{Conclusion and Future Directions}

Our MI-based auditory nerve modeling framework identified 40\%-compressed VCV syllables in AM noise as the condition producing the largest modeled information loss. This was also the only tested condition that significantly distinguished normal-audiogram listeners with and without self-reported hearing difficulty, providing support for using the framework to guide stimulus design. However, the behavioural differences cannot yet be attributed specifically to CND alone, and individual SRT$_{50}$ calibration minimizes cognitive and central confounds but does not fully separate peripheral from central contributions to performance.  Establishing the sensitivity and specificity of these probes to CND remains a focus of future work. We will incorporate physiological testing, including envelope-following responses (EFRs), ABR wave~I, and middle-ear muscle reflex measurements, to examine whether the behavioural compression penalty is associated with physiological indicators of cochlear neural dysfunction as well.

\section{Generative AI Disclosure}
Generative AI was used only to assist with editing and polishing the manuscript text. All scientific content, experimental design, code, data analysis, interpretation of results and intellectual contributions are solely the work of the authors, who reviewed and take full responsibility for the final content.


\ninept
\bibliographystyle{IEEEbib}
\bibliography{strings,mybib}

\end{document}